\documentclass[12pt]{article}
\usepackage{epsf}

\def\({\c c}
\def\|{\'\i}

\def\nl {\par \noindent }

\newcommand{\VEV}[1]{\left\langle{#1}\right\rangle}

\newcommand{\smallbox}{\fbox{\rule[.91mm ]{0cm}{.91mm}\,\ }}

\begin {document}

\begin{flushright}
{\small
SLAC--PUB--8711\\
November 2000\\
revised April 2001}
\end{flushright}
\medskip

\begin{center}
{\large\bf LIGHT-FRONT-QUANTIZED QCD IN LIGHT-CONE GAUGE:
THE DOUBLY TRANSVERSE GAUGE PROPAGATOR}
\footnote{Research partially supported by the Department of Energy under
contract DE-AC03-76SF00515 }

\vspace{1cm}

{\large Prem P. Srivastava}\footnote{E-mail:\quad prem@uerj.br
or prem@cbpf.br}

\vspace{0.3cm}
{\em \it Instituto de F\'{\i}sica, UERJ-Universidade do
Estado de Rio de Janeiro, Brasil}

\smallskip

and

\smallskip

{\large Stanley J. Brodsky}\footnote{E-mail:\quad sjbth@slac.stanford.edu}

\vspace{0.3cm}
{\em \it Stanford Linear Accelerator Center \\
 Stanford University,
Stanford, California 94309}
\vspace{0.2cm}
\end{center}

\vfill \eject

\begin{center}
{\bf Abstract}
\end{center}

The light-front (LF) quantization of QCD in light-cone gauge has a
number of remarkable advantages, including explicit unitarity, a
physical Fock expansion, the absence of ghost degrees of freedom, and
the decoupling properties needed to prove factorization theorems in high
momentum transfer inclusive and exclusive reactions.  We present a
systematic study of LF-quantized gauge theory following the Dirac method
and construct the Dyson-Wick S-matrix expansion based on LF-time-ordered
products.
The free theory gauge field is shown to satisfy the Lorentz condition as
an operator equation as well as the light-cone gauge condition.  Its
propagator is found to be transverse with respect to both its
four-momentum and the gauge direction.
The interaction Hamiltonian of QCD can
be expressed in a form resembling that of covariant theory, except for
additional instantaneous interactions which can be treated
systematically.
 The renormalization constants in YM theory are shown to
satisfy the identity $Z_1=Z_3$ at one loop order.
The QCD $\beta$ function computed in the noncovariant light-cone
gauge agrees with that known in the conventional framework.
 Some comments on the relationship of our LF framework, with the
doubly transverse gauge propagator, to the
analytic effective charge and renormalization scheme defined by the
pinch technique, the unitarity relations and the spectral representation
 are also made.  LF quantization thus provides a consistent
formulation of gauge theory, despite the fact that the hyperplanes
$x^{\pm}=0$ used to impose boundary conditions constitute characteristic
surfaces of a hyperbolic partial differential equation.

\vfill

\begin{center}
 to appear in Physical Review D15, July 2001
\end{center}
\nl {\bf Keywords}:\quad{Quantum Chromodynamics; Quantization;
Light-cone gauge; Light-front; Constraints; Gauge Theory;  Propagator}
\vfill
\newpage

\section{Introduction}
\label{intrO}

The quantization of relativistic field theory at fixed light-front time
$\tau = (t - z/c)/\sqrt 2$, which was proposed by Dirac \cite{dir} half
a century ago, has found important applications \cite{bro,bpp,ken,pre}
in both gauge theory and string theory \cite{{susskind}}.  The
light-front (LF) quantization of QCD in its Hamiltonian form provides an
alternative to lattice gauge theory for the computation of
nonperturbative quantities such as the spectrum and the light-front Fock
state wavefunctions of relativistic bound states \cite{bpp}.  LF
variables have also found natural applications in other contexts, such
as in the quantization of (super-) string theory and M-theory
\cite{susskind}.  Light-front quantization has been employed in the
nonabelian bosonization \cite{wit} of the field theory of $N$ free
Majorana fermions and was used in the demonstration of the asymptotic
freedom of the Yang-Mills theory beta-function \cite{thorn}.  The
requirement of the microcausality\cite{weif} implies that the LF
framework is more appropriate for quantizing \cite{prec} the self-dual
(chiral boson) scalar field.

Since LF
coordinates are not related to the conventional coordinates by a finite
Lorentz transformation, the descriptions of the same physical result may
be different in the equal-time ({\it instant form}) and equal-LF-time
({\it front form}) formulations of the theory.  This was in fact found
to be the case in a recent study \cite{pre1, pre1c} of some soluble
two-dimensional gauge theory models, where it was also demonstrated that
LF quantization is very economical in displaying the relevant degrees of
freedom, leading directly to a physical Hilbert space.
The corresponding Fock representation is boost independent since the
{\it front form} has seven kinematical Poincar\'e generators \cite{dir},
including Lorentz boost transformations, compared to only six in the
{\it instant form} framework.
LF-time-ordered perturbation theory is much more economical than
equal-time-ordered perturbation theory, since only graphs with particles
with positive LF momenta $p^+ = (p^0 + p^3)/\sqrt 2$ appear.
LF-time-ordered perturbation theory has also been applied \cite
{soper,yan} to massive fields.  It was used in the analysis of the
evolution of deep inelastic structure functions \cite{Bjorken:1971ah} and
the evolution of the distribution amplitudes which control hard exclusive
processes in QCD \cite{Lepage:1980fj}. 
Recently, light-cone gauge on the light-front has been used to analyze the forces between
fixed colored sources \cite{Rozowsky:1999pa}, the string structure of
QCD at large $N_C$ \cite{Bering:2000cw}, and spontaneous symmetry breaking phenomena without
zero modes \cite{Rozowsky:2000gy}.  There have also been interesting applications to
supersymmetric theories on the light-front \cite{Pinsky:2000rn}.

It has been conventional to apply LF quantization to gauge theory in
light-cone (l.c.) gauge $A^+=A_{-} = (A^0 + A^3)/\sqrt 2=0$, since the
transverse degrees of freedom of the gauge field can be immediately
identified as the dynamical degrees of freedom, and ghost fields can be
ignored in the quantum action of the nonabelian gauge theory
\cite{Lepage:1980fj, Bassetto:1997ba, leibb}.  The light-front (LF)
quantization of quantum chromodynamics in l.c. gauge thus has a number
of remarkable advantages, including explicit unitarity, a physical Fock
expansion, and the complete absence of ghost degrees of freedom.
In addition, the decoupling of gluons to propagators carrying high momenta
and the absence of
collinear divergences in irreducible
diagrams in the l.c. gauge
are important tools for proving the leading-twist factorization of soft
and hard gluonic corrections in high momentum transfer inclusive and
exclusive reactions \cite{Lepage:1980fj}.
On the negative side, any noncovariant gauge brings
in the breaking of manifest rotational invariance, instantaneous
interactions, and, apparently, a more difficult renormalization procedure
\cite{ Bassetto:1997ba,leibb,{Perry:1999sh}}.

In this paper we will discuss the LF quantization of QCD gauge field
theory in l.c. gauge employing the Dyson-Wick S-matrix expansion
\cite{bjk} based on LF-time-ordered products \cite{ref:conf}.
We shall first study the gauge-fixed quantum action of the theory on the LF.
The LF Hamiltonian framework will then be constructed following the
Dirac method \cite{dir1,wei1} which allows one to self-consistently
identify the independent fields and their commutation relations
in the presence of the l.c. gauge condition and other constraints.
It also allows us to study \cite{dir1}
the Lorentz covariance properties of the theory.

The LF framework is a severely constrained dynamical theory
with many second-class constraints.  These can be eliminated by
constructing Dirac brackets, and the theory can be quantized canonically
by the correspondence principle in terms of a {\it reduced number} of
independent fields.  The commutation relations among the field operators
are also found by the Dirac method, and they are used to obtain the
momentum space expansions of the fields.  For example, the nondynamical
projection of the fermion field can be eliminated using a nonlocal
constraint equation.
The gauge-field quantization of the {\it massless} field
in the l.c. gauge in the {\it front form } theory is studied.
Using the derived commutators, we find that LF
quantized free gauge theory simultaneously satisfies the covariant gauge
condition $\partial\cdot A=0$ as an operator condition as well as the
light-cone gauge condition.  The Fourier transform of the free theory
gauge field and its propagator in momentum space then
follow straightforwardly.
The removal
of the unphysical components of the fields results in \cite{bro,
cov} tree-level instantaneous interaction terms which can be evaluated
systematically (See Sections 4 and 5).  The instantaneous light-cone
gauge interactions of the light-front Hamiltonian are
incorporated into nonperturbative approaches such as DLCQ
\cite{Pauli:1985ps}.

The QCD interaction Hamiltonian is constructed in Section 4 where we
restore in the expression the dependent components $A_{+}$ and
$\psi_{-}$.  It then takes a form close to that of covariant gauge
theory without ghost terms, plus instantaneous interactions which are
straightforward to handle in the Dyson-Wick perturbation theory.

The renormalization procedure in our framework is illustrated in Section
5 by considering the nonabelian YM gauge theory.
The equality $Z_{1}=Z_{3}$ is explicitly demonstrated to one loop
in our doubly transverse gauge framework
and the $\beta-$function is found to agree with that known in
the conventional Feynman gauge calculation.
 The results are compared with those found
\cite{Bassetto:1997ba, leibb, schweda} in the conventional l.c. gauge
equal-time framework.
Gluon self-energy coming from quark loops is
also computed.
A calculation of the electron-muon scattering amplitude in QED is used to
show the relevance of instantaneous interactions for
recovering the Lorentz invariance.
We recall that
the Dyson-Wick expansion has been used \cite{pre2} to renormalize
two-dimensional scalar field theory on the LF with nonlocal
interactions.
 Appendix C comments on the understanding in the gluon self-energy
of a noncovariant $\log$ term, which arises from another basic
noncovariant divergent integral, present in the noncovariant gauge under
study.  Its relevance in the context of
 the spectral representation and unitarity is briefly touched upon.
The complete renormalization of QCD in our framework,
 including the
verification of Slavnov-Taylor identities,  will
be considered in a forthcoming paper.

\section{QCD Action in Light-Cone Gauge}

The LF coordinates are defined as
$x^{\mu}=(x^{+}=x_{-}=(x^{0}+x^{3})/{\sqrt 2},\, x^{-}=x_{+}=
(x^{0}-x^{3})/{\sqrt 2},\, x^{\perp})$, where $x^{\perp}= (x^{1},x^{2})
=(-x_{1},-x_{2})$ are the transverse coordinates and $\,\mu=-,+,1,2$.
The coordinate $x^{+}\equiv \tau$ will be taken as the LF time, while
$x^{-}$ is the longitudinal spatial coordinate.  We can of course choose
a convention where the role of $x^{+}$ and $x^{-}$ interchanged.  The
equal-$x^{+}$ quantized theory already contains the information on the
equal-$x^{-}$ commutator \cite{pre1,pre1c}.  The LF components of any
tensor, for example, the gauge field, are similarly defined, and the
metric tensor $g_{{\mu}{\nu}}$ may be read from $A^{\mu} B_{\mu}=
A^{+}B^{-}+A^{-}B^{+} -A^{\perp}B^{\perp}$.  Also $k^{+} $ indicates the
longitudinal momentum, while $k^{-}$is the corresponding LF energy.

The quantum action of QCD in l.c. gauge is described in standard
notation by the following Lagrangian density
\begin{equation} {\cal
L}_{QCD}=-{1\over 4}F^{a\mu\nu}{F^{a}}_{\mu\nu}+ B^{a} {A^{a}}_{-} +
{\bar c}^{a}{\cal D}^{ab}_{-} c^{b} + {\bar\psi}^{i}
(i\gamma^{\mu}{D^{ij}}_{\mu}-m\delta^{ij})\psi^{j}
\end{equation}
Here
$\psi^{j}$ is the quark field with color index $j=1..N_{c}$ for an
$SU(N_{c})$ color group, ${A^{a}}_{\mu}$ the gluon field,
$F^{a}_{\mu\nu}=\partial_{\mu}A^{a}_{\nu} -\partial_{\nu}{A^{a}}_{\mu}
+g f^{abc}{A^{b}}_{\mu}{A^{c}}_{\nu}$ the field strength, ${{\cal
D}^{ac}}_{\mu}= (\delta^{ac}\partial_{\mu}+g f^{abc}{A^{b}}_{\mu})$,
${D^{ij}}_{\mu}\psi^{j}= (\delta^{ij}\partial_{\mu}-ig {A^{a}}_{\mu}
{t^{a}}^{ij})\psi^{j} $, $t^{a}\equiv \lambda^{a}/2$,
$a=1..({N_{c}}^{2}-1)$, the gauge group index, and $\,{\bar c}^{a},
c^{a}$ are anti-commuting ghost fields.  In writing the quantum action
we introduce auxiliary Lagrange multiplier fields $B^{a}(x)$ and add to
the Lagrangian the {\it linear} gauge-fixing term $(B^{a}{A^{a}}_{-})$,
which is a traditional procedure.  In addition we are required to
also add ghost terms such that the action (1) becomes invariant under BRS
symmetry \cite{stora} transformations.

It is worth recalling the corresponding procedure for implementing a
covariant gauge-fixing condition.  For example, in Feynman
gauge one adds the term
$(B^{a}\partial^{\mu} A_{a\mu} + B^{a}B^{a}/2)$ to the Lagrangian.
The quadratic $B^{a}B^{a}$ term is allowed on dimensional
considerations.  However in the case of l.c. gauge, the auxiliary field
$B^{a}$ carries canonical dimension three and as such a quadratic term is
not allowed in (1).  We mention yet another example:  the
quantum action for constructing \cite{prec} the quantized theory of the
self-dual scalar field (chiral boson) in two-dimensional space-time.  One
starts by adding the traditional linear term $ B\partial_{-}\phi$ to the
free scalar field Lagrangian.  Its LF quantization can be performed
without any violation of the principle of microcausality, in contrast to
that occurs in the conventional treatment.  The quantized theory is
found to be trivial indicating that the traditional Lagrange multiplier
method breaks down at the quantum level.  However, if we add to
the theory an additional $ B^{2}$ term, which is allowed on dimensional
considerations, the LF quantization of the improved theory does produce
a satisfactory description \cite{prec} of the quantized
left- and right-movers.  The theory also contains as a special case the
well known Floreanini-Jackiw action \cite{flor}, giving a plausible
reason for the success of that model.

The quark field term in LF coordinates reads in the notation of Appendix B as
\begin{eqnarray}
{\bar\psi}^{i} (i\gamma^{\mu}{D^{ij}}_{\mu}-m\delta^{ij})\psi^{j} &=& i{
\sqrt {2}}{\bar\psi_{+}}^{i}
\gamma^{0}{D^{ij}}_{+}{\psi_{+}}^{j}+{\bar\psi_{+}}^{i}
(i\gamma^{\perp}{D^{ij}}_{\perp}-m\delta^{ij}){\psi_{-}}^{j} \nonumber
\\ &+&{\bar\psi_{-}}^{i}\left[{i \sqrt {2}}\gamma^{0}
{D^{ij}}_{-}{\psi_{-}}^{j}+
(i\gamma^{\perp}{D^{ij}}_{\perp}-m\delta^{ij}) {\psi_{+}}^{j}\right]
\end{eqnarray}
This shows that the minus components ${\psi_{-}}^{j}$ are in fact
nondynamical (Lagrange multiplier) fields without kinetic terms.
The variation of the action with respect to ${{\bar{\psi^{j}}}_{-}}$ and
${{{\psi^{j}}}_{-}}$ leads to the following gauge-covariant constraint
equation
\begin{equation}
i{\sqrt 2}\,\, {D^{ij}}_{-}{\psi_{-}}^{j}=
-(i\gamma^{0}\gamma^{\perp}{D^{ij}}_{ \perp}-m\gamma^{0}\delta^{ij})
{\psi_{+}}^{j},
\end{equation}
and its conjugate.  The ${\psi^{j}}_{-}$
components may thus be eliminated in favor of the independent
dynamical component
$\psi_{+}^{j}$.  It
gives rise to instantaneous terms in the interaction Hamiltonian
given in Sec 4 and also the free theory propagator of $\psi_{+}$ is
found \cite{cov} to be
 causal and carries no instantaneous term.

\section{Gauge Field Propagator in l.c. Gauge}

The quadratic terms in the Lagrangian density which determine
the free gauge field propagators are
\begin{equation}
\frac{1}{2}\left[{F^{a}}_{+-}{F^{a}}_{+-}
+2{F^{a}}_{+{\perp}}{F^{a}}_{-{\perp}} -{F^{a}}_{12}{F^{a}}_{12} \right]
+ B^{a}{A^{a}}_{-}+ {\bar c}^{a}\partial_{-}c^{a}.
\end{equation}
We
observe that in the {\it front form} framework, the fields ${A^{a}}_{+}$
as well as $B^{a}$ have no kinetic terms, and they enter in the action as
auxiliary multiplier fields.  Also, since the ghost fields decouple, it
is sufficient to study the free abelian gauge theory with the following
action
\begin{equation}
\int d^{2}x^{\perp}dx^{-} \left\{{1\over
2}\left[(F_{+-})^{2}- (F_{12})^2 +2F_{+\perp} F_{-\perp}\right]+ B A_{-}
\right\}
\end{equation}
where $F_{\mu\nu}$ stands for
$(\partial_{\mu}A_{\nu}-\partial_{\nu}A_{\mu})$ in the present section.
The gauge field equations of motion are $\,\smallbox A_{\mu}=
\partial_{\mu}(\partial\cdot A) - B {\delta_{\mu}}^{-}$, $A_{-}=0$,
$\mu=-,+,\perp\,$ and $\perp=1,2$, and as a consequence
$\partial_{-}B=0$.  The canonical momenta following from (5) are
$\pi^{+}=0$, $\pi_{B}=0$, $\pi^{\perp}=F_{-\perp}$, and $\pi^{-}=F_{+-}=
(\partial_{+} A_{-}-\partial_{-} A_{+})\,$ which indicates that we are
dealing with a constrained dynamical system.  The Dirac procedure
 will be followed in order to construct the self-consistent
Hamiltonian theory which is required for performing canonical
quantization.  The canonical
Hamiltonian density is
\begin{equation}
{\cal H}_{c}= {1\over 2}
({\pi}^{-})^{2}+{1\over 2}(F_{12})^{2}-
A_{+}(\partial_{-}\pi^{-}+\partial_{\perp}\pi^{\perp})- B A_{-}
\end{equation}
The primary constraints following from (5) are $\pi^{+}\approx 0$,
$\,\pi_{B}\approx 0$ and
$\,{\eta^{\perp}}\equiv\pi^{\perp}-\partial_{-}A_{\perp}+
\partial_{\perp}A_{-}\approx 0$, where $\,\approx\,$ stands for the {\it
weak equality} relation.  We now require the persistency in
$\tau$ of these constraints employing the preliminary Hamiltonian, which
is obtained by adding to the canonical Hamiltonian the primary
constraints multiplied by the undetermined Lagrange multiplier fields
$u_{+}$, $u_{\perp}$, and $u_{B}$.  In order to obtain the Hamilton's
equations of motion, we assume initially the standard Poisson brackets
for all the dynamical variables present in (6).

We are then led to the following secondary constraints
\begin{eqnarray}
\Phi\equiv \partial_{-}\pi^{-}+\partial_{\perp}\pi^{\perp} & \approx &
0, \nonumber \\ A_{-} & \approx & 0
\end{eqnarray}
which are already
present in (6) multiplied by Lagrange multiplier fields.  Requiring
also the persistency of $\Phi$ and $A_{-}$ leads to another secondary
constraint
\begin{equation}
\Psi\equiv\pi^{-}+\partial_{-}A_{+}\approx 0.  \
\end{equation}
The procedure stops at this stage, and no more
constraints are seen to arise since further repetition leads to
equations which would merely determine the multiplier fields.

Let us now analyze the
nature of the phase space constraints.  In
spite of the gauge-fixing term introduced in the initial Lagrangian,
there still remains on the canonical LF phase space a first class
constraint $\pi_{B}\approx 0$.  An inspection of the
equations of motion shows that we may add \cite{dir1} to the set of constraints
found above an additional external constraint $B\approx
0$.  This would make the whole set of constraints in the theory
second class.
Dirac brackets satisfy the property such that
we can set the above set of constraints as {\it strong} equality
relations inside them.
The equal-$\tau$ Dirac
bracket $\{f(x),g(y)\}_{D}$ which carries this property is
straightforward to construct.  Hamilton's equations now employ
 the Dirac brackets rather than the Poisson ones.
The phase space constraints on the light front:
$\pi^{+}= 0$, $\eta^{\perp}= 0$, $A_{-}=0$, $\Phi=0$, $\Psi= 0$,
$\pi_{B}=0$, and $B=0$ thus effectively eliminate $B$ and
all the canonical momenta
from the theory.  The surviving dynamical variables are $A_{\perp}$
while $A_{+}$ is a dependent variable which satisfies
$\partial_{-}(\partial_{-}A_{+}-\partial_{\perp}A_{\perp})=0$.  The {\it
reduced} Hamiltonian is found to be
\begin{equation}\label{eq:h0}
{H_{0}}^{LF} = {1\over 2}\int {d^{2}x^{\perp}}dx^{-}\;
\left[(\partial_{-}A_{+})^{2} + \frac{1}{2}
F_{\perp\perp'}F^{\perp\perp'} \right],
\end{equation}
where we have retained the dependent variable
$A_{+}$ for convenience.

The canonical quantization of the theory at equal-$\tau$ is performed
 {\it via} the correspondence $i\{f(x),g(y)\}_{D} \to
\left[f(x),g(y)\right]$ where the latter indicates the commutators among
the corresponding field operators.  The equal-LF-time commutators of the
transverse components of the gauge field are found to be
$$\left[A_{\perp}(\tau,x^{-},x^{\perp}),A_{\perp'}
  (\tau,y^{-},y^{\perp})\right]
     =i\delta_{\perp\perp'} K(x,y)$$
where
$K(x,y)=-(1/4)\epsilon(x^{-}-y^{-})\delta^{2}(x^{\perp}-y^{\perp})$.
The commutators are nonlocal in the longitudinal coordinate but there is
no violation \cite{weif} of the microcausality principle on the LF.
 At equal LF-time, $(x-y)^2 = -(x^{\perp}-y^{\perp})^{2} <0 $, is
nonvanishing for
$x^{\perp}\neq y^{\perp}$ but $\delta^{2}(x^{\perp}-y^{\perp})$
vanishes for such spacelike separation.
The commutators of the transverse components of the gauge fields are
physical, having the same form as the commutators of scalar fields in
the {\it front form} theory.

The Heisenberg equations of motion employing (9) lead to the Lagrange
equations for the independent fields which assures us of the
self-consistency \cite{dir1} of the {\it front form} Hamiltonian theory
in the l.c. gauge.  We also find that the
commutators of $A_{+}$ are identical to the ones obtained by
substituting $A_{+}$ by $(\partial_{\perp}/ \partial_{-}) A_{\perp}$.
This is a consequence of the definition of the Dirac bracket itself and
manipulations on it with the partial derivatives.
Hence in the free l.c.
gauge theory on the LF we obtain the Lorentz condition $\partial\cdot A=0$ as
an operator equation as well.
The LF commutators of the gauge field may be realized in
momentum space by the following Fourier transform
\begin{equation}\label{eq:  gaugefield}
A_{\perp}(x)={1\over {\sqrt {(2\pi)^{3}}}} \int d^{2}k^{\perp}dk^{+}\,
{\theta(k^{+})\over {\sqrt {2k^{+}}}}\, \left[a_{\perp}(\tau,
k^{+},k^{\perp}) e^{-i{\bar k}\cdot{x}} +a^{\dag}_{\perp}(\tau,
k^{+},k^{\perp}) e^{i{ \bar k}\cdot{ x}} \right ]
\end{equation}
where
${ \bar k}\cdot{ x}= k^{+}x^{-}-k^{\perp}x^{\perp}\,$ and $a_{\perp}$, $
a^{\dag}_{\perp}$ are operators which satisfy the equal-$\tau$ canonical
commutation relations with the nonvanishing ones given by
$\,[a_{\perp}(\tau,k^{+}, k^{\perp}),$
${a^{\dag}}_{\perp'}(\tau,k'^{+},k'^{\perp}) ]$ $=\delta_{\perp\perp'}$
$\delta^{3}(k-k')$ where $\, \delta^{3}(k-k')\equiv\delta(k^{+}-k'^{+})$
$\delta^{2}(k^{\perp}-k^{\perp'})$.  The Heisenberg equation of motion
for $A_{\perp}(x)$ then leads to $\,a_{\perp}(\tau, k^{+},k^{\perp})=$ $
a_{\perp}(k^{+},k^{\perp})$ $exp (-ik^{-}x^{+})$ where $k^{-}$ is
defined through the dispersion relation, $\, 2 k^{-}k ^{+}=
k^{\perp}k^{\perp}$.  The operators $a_{\perp}(k^{+},k^{\perp})$ and
${a^{\dag}}_{\perp}(k^{+},k^{\perp})$ are thus associated with the {\it
massless} gauge field quanta.  The Fourier transform (10) may then be
rewritten as
\begin{equation}\label{eq:  field}
A_{\perp}(x)={1\over
{\sqrt {(2\pi)^{3}}}} \int d^{2}k^{\perp}dk^{+}\, {\theta(k^{+})\over
{\sqrt {2k^{+}}}}\, \left[a_{\perp}( k^{+},k^{\perp}) e^{-i{ k}\cdot{x}}
+a^{\dag}_{\perp}(k^{+},k^{\perp}) e^{i{ k}\cdot{ x}} \right ]
\end{equation}
where $k\cdot x
=(k^{-}x^{+}+k^{+}x^{-}+k_{\perp}x^{\perp})$ and $k^{\mu}k_{\mu}=0$.
The Fourier transform (11) is of the typical form of the {\it front
form} theory where the bosonic fields satisfy nonlocal LF commutation
relation; it does not carry in it any explicit information on the mass
of the field.
The commutators of $A_{+}$ are realized if we write
for its Fourier transform
\begin{equation}\label{eq:  field1}
A_{+}(x)={1\over {\sqrt {(2\pi)^{3}}}} \int d^{2}k^{\perp}dk^{+}\,
{\theta(k^{+})\over {\sqrt {2k^{+}}}}\, \left[a_{+}( k^{+},k^{\perp})
e^{-i{ k}\cdot{x}} +a^{\dag}_{+}(k^{+},k^{\perp}) e^{i{ k}\cdot{ x}}
\right ]
\end{equation}
where $\,a_{+}(k)\,$ is determined from
$\,[\,k^{+}a_{+}(k)+ k^{\perp}a_{\perp}(k)\,]=0$.

The free
propagators in momentum space are derived straightforwardly.  We find
\begin{eqnarray}
\VEV{0|\,T({A^{a}}_{\perp}(x){A^{b}}_{\perp}(0))\,|0} &=&
\VEV{0|\left[\theta(\tau){A^{a}}_{\perp}(x){A^{b}}_{\perp}(0)
+\theta(-\tau) {A^{b}}_{\perp}(0){A^{a}}_{\perp}(x)\right]|0} \nonumber
\\[1ex] &=& {{i\delta^{ab}} \over {(2\pi)^{4}}} \int d^{4}k
\;e^{-ik\cdot x}\; \; {-g_{\perp\perp'}\over {k^{2}+i\epsilon}}
\end{eqnarray}
where we have restored the gauge index $a$.  In view of
(12) we may write the gauge field propagator in the l.c. gauge in the
following convenient form
\begin{equation}
\VEV{0|\,T({A^{a}}_{\mu}(x){A^{b}}_{\nu}(0))\,|0} ={{i\delta^{ab}} \over
{(2\pi)^{4}}} \int d^{4}k \;e^{-ik\cdot x} \; \; {D_{\mu\nu}(k)\over
{k^{2}+i\epsilon}}
\end{equation}
where we have defined
\begin{equation}
D_{\mu\nu}(k)= D_{\nu\mu}(k)= -g_{\mu\nu} + \frac
{n_{\mu}k_{\nu}+n_{\nu}k_{\mu}}{(n\cdot k)} - \frac {k^{2}} {(n\cdot
k)^{2}} \, n_{\mu}n_{\nu}.
\end{equation}
Here $n_{\mu}$ is a null
four-vector, gauge direction, whose components are chosen to be $\,
n_{\mu}={\delta_{\mu}}^{+}$, $\, n^{\mu}={\delta^{\mu}}_{-}$.  We note that
\begin{eqnarray}
D_{\mu\lambda}(k) {D^{\lambda}}_{\nu}(k)=
D_{\mu\perp}(k) {D^{\perp}}_{\nu}(k)&=& - D_{\mu\nu}(k), \nonumber \\
k^{\mu}D_{\mu\nu}(k)=0, \qquad \quad && n^{\mu}D_{\mu\nu}(k)\equiv
D_{-\nu}(k)=0, \nonumber \\ D_{\lambda\mu}(q) \,D^{\mu\nu}(k)\,
D_{\nu\rho}(q') &=& -D_{\lambda\mu}(q)D^{\mu}_{\rho}(q').
\end{eqnarray}
The property that
the gauge field propagator $\,\,i\,D_{\mu\nu}(k)/ (k^{2}+i\epsilon)\,$ is
transverse not only to the gauge direction $n_{\mu}$ but also to
$k_{\mu}$, {\em i.e.}, it is doubly transverse, leads to appreciable
simplifications in the computations in QCD as is illustrated below.  In
a sense our gauge propagator corresponds to the form used in Landau gauge,
but here it is derived in the context of the noncovariant l.c. gauge.  As
usual with the non-covariant gauges, the propagator contains a
non-covariant piece added to the covariant (Feynman gauge) propagator.
It differs from the propagators derived \cite{Bassetto:1997ba,leibb,schweda}
in equal-time quantized l.c. gauge QCD.
The form (14) of the propagator reminds us of
the rules, in the context of the old-fashioned perturbation theory, laid
down in Ref. \cite{Lepage:1980fj} a long time ago, in the context of LF
quantization.  We will comment in Section 5 on the problem of handling the
singularity near $(n\cdot k) \approx 0$ present in the propagator.

We can introduce the operators $\,b_{(\perp)}$ and
${b^{\dag}}_{(\perp)}$, $(\perp)=(1), (2)$ representing the two
independent states of transverse polarizations of a
massless photon.  They are assumed to obey the standard canonical
commutation relations
$\,[b_{(\perp)}(k^{+},k^{\perp}),{b^{\dag}}_{(\perp')}(k'^{+},k'^{\perp})
]$ $=\delta_{(\perp)(\perp')}$ $\delta^{3}(k-k')$.  We write
$\,a^{\perp}(k)= \sum_{(\perp')}$ ${E^{\perp}}_{ (\perp')}(k) $
$b_{(\perp')}(k)$ where the $\,{E_{(\perp)}}^{\mu}(k)\,$ indicate the two
independent polarization four-vectors.  A convenient set may be chosen
to be
\begin{equation}
{E_{(\perp)}}^{\mu}(k)= E^{(\perp){\mu}}(k)= -
\,D^{\mu}_{\perp} (k) \end{equation} which have the property
\begin{eqnarray} \sum_{{(\perp)}=1,2}
{E^{(\perp)}}_{\mu}(k){E^{(\perp)}}_{\nu}(k)= D_{\mu\nu}(k) , &&\qquad
\quad g^{\mu\nu} {E^{(\perp)}}_{\mu}(k){E^{(\perp')}}_{\nu}(k)= g^{\perp
\perp'} \\ k^{\mu}E^{(\perp)}_{\mu}(k)=0, \qquad && \qquad \quad n^{\mu}
E^{(\perp)}_{\mu} \equiv E^{(\perp)}_{-}=0
\end{eqnarray}
The Fourier
transform of the gauge field may then be expressed in the standard form
\begin{equation}
A^{\mu a}(x)={1\over {\sqrt {(2\pi)^{3}}}} \int
d^{2}k^{\perp}dk^{+}\, {\theta(k^{+})\over {\sqrt {2k^{+}}}}\,
\sum_{(\perp)} {E_{(\perp)}}^{\mu}(k) \left[b_{(\perp)}^{a}(
k^{+},k^{\perp}) e^{-i{ k}\cdot{x}} +b^{\dag a}_{(\perp)}(k^{+},k^{\perp})
e^{i{ k}\cdot{ x}} \right ]
\end{equation}
where the l.c. gauge
$\,A_{-}^{a}=0$, along with the Lorentz condition, is already incorporated
in it.  The
momentum space expressions of LF energy and momentum confirm the
interpretation of $b_{(\perp)}$ and ${b^{\dag}}_{(\perp)}$ as the Fock
space operators of annihilation and creation of massless transverse
gauge field quanta.  Only the physical transverse degrees of freedom
appear in the gauge field expansion.

\section {The QCD Hamiltonian in l.c. Gauge}

The Dyson-Wick perturbation theory expansion in the interaction
representation requires that we separate the full Hamiltonian into
the free theory component and
 coupling-constant-dependent interaction piece.

The equations of motion in LF coordinates following from (1) give
\begin{equation}
2i
\partial_{-}{\psi^{i}}_{-}=2i \partial_{-} {{\widetilde\psi}^{i}}_{-} +g
\gamma^{\perp} {A_{\perp}}^{a} (t^{a})^{ij} \gamma^{+} {\psi^{j}}_{+}
\end{equation} and \begin{equation} 2i
\partial_{+}{\psi^{i}}_{+}=
(i\gamma^{\perp}\partial_{\perp}+m)\gamma^{-} {\psi^{i}}_{-}
+g \gamma^{\perp} {A_{\perp}}^{a} (t^{a})^{ij} \gamma^{-} {\psi^{j}}_{-}
-2g {A_{+}}^{a} (t^{a})^{ij} {\psi^{j}}_{+},
\end{equation}
along with
\begin{equation}
\partial_{-}(\partial_{-}{A_{+}}^{a}-\partial_{-}{{\widetilde A}_{+}}^{a})
=-g f_{abc} {A_{\perp}}^{b} \partial_{-}{A^{\perp}}^{c} +g
{\bar\psi}^{i} \gamma^{+} (t^{a})^{ij} \psi^{j}
\end{equation}
where we
define \cite {bro} ${{\widetilde A}_{+} }^{a}$ and
${{\widetilde\psi}^{i}}_{-}$
by $\partial_{-}{{\widetilde A}_{+}}^{a}=\partial_{\perp}{A_{\perp}}^{a}$
and $ 2i \partial_{-}{{\widetilde\psi}^{i}}_{-}=
(i\gamma^{\perp}\partial_{\perp}+m)\gamma^{+} {\psi^{i}}_{+}$ respectively.
 The combination $({\psi^{i}}_{+} + {{\widetilde\psi}^{i}}_{-})$, when $g=0$,
satisfies the free Dirac equation.
Hence the interaction Hamiltonian in the l.c. gauge, $A^{a}_{-}=0$,
can be rewritten\footnote{We note that the dependent field $\psi_{-}$ and
$A_{+}$ occur only in the first two terms of (24).}
in the following useful form: \cite{bro}
\begin{eqnarray}
{\cal H}_{int}= -{\cal
L}_{int}&=& -g \,{{\bar\psi}}^{i} \gamma^{\mu}\, (t^{a})^{ij}\,
{{\psi}}^{j} \,A_{\mu}^{a}
\nonumber \\ && +\frac{g}{2}\, f^{abc} \,(\partial_{\mu}{A^{a}}_{\nu}-
\partial_{\nu}{A^{a}}_{\mu}) A^{b\mu} A^{c\nu} \nonumber \\ && +\frac
{g^2}{4}\, f^{abc}f^{ade} {A_{b\mu}} {A^{d\mu}} A_{c\nu} A^{e\nu}
\nonumber \\ && - \frac{g^{2}}{ 2}\,\, {{\bar\psi}}^{i} \gamma^{+}
\,\gamma^{\mu}\,A_{\mu}^{a}\,(t^{a})^{ij}\,\frac{1}{i\partial_{-}} \,
\gamma^{\nu} \,A_{\nu}^{b}\,(t^{b})^{jk}\,{\psi}^{k}
 \nonumber \\
&& -\frac{g^{2}}{ 2}\, (\frac{1}{i\partial_{-}}\,j^{+}_{a})\,
(\frac{1}{i\partial_{-}}\,j^{+}_{a} )
\end{eqnarray}
where
\begin{equation}
j^{+}_{a}={{\bar\psi}}^{i} \gamma^{+} ( {t_{a}})^{ij}{{\psi}}^{j} +
f_{abc} (\partial_{-} A_{b\mu}) A^{c\mu}
\end{equation}
and a sum over distinct quark and lepton flavors (in QED), not
written explicitly, is understood in (25) and (24).

The perturbation theory expansion in the interaction representation,
where we time-order with respect to the LF time $\tau$, can now be built
following the Dyson-Wick \cite {bjk} procedure.  There are no ghost
interaction terms to consider.  The instantaneous interaction
contributions (the last two terms in (24)) can be dealt with
systematically.  Such terms are also required in abelian QED
theory, obtained by suppressing in the above interaction the
additional terms of nonabelian theory.  For example,  the tree level {\it
seagull} term dominates the classical Thomson formulae for the
scattering at the vanishingly small photon energies.  The instantaneous
counterterms also serve to
restore the manifest Lorentz invariance,
which was broken by the use of noncovariant l.c. gauge and the
noncovariant propagator.
The information on the l.c. gauge is encoded in the remarkable
properties of the gauge field propagator in the LF framework.
Some of the vertices in momentum space required for the illustrations
below are summarized in the Appendix A.

\section {Illustrations}
\subsection{Electron-Muon Scattering}

The contributions to the matrix element from the mediation of the gauge
field is
\begin{equation}
-e^{2}\left[{\bar u}_{e
}(p'_{1})\gamma^{\mu}{u_{e }(p_{1})}\; {\bar u}_{\mu
}(p'_{2})\gamma^{\nu} {u_{\mu }(p_{2})}\right]\;\; {{i\,D_{\mu
\nu}}(q)\over {q^{2}+i\epsilon}}.
\end{equation}
where
$q=-p'_{1}+p_{1}=p'_{2}-p_{2}$.  Using the mass-shell conditions for the
external lines it reduces to
\begin{equation}
-i\,e^{2}\left[{\bar u}_{e}(p'_{1})\gamma^{\mu}{u_{e }(p_{1})}\;
{\bar u}_{{\mu}}
(p'_{2})\gamma^{\nu} {u_{{\mu} }(p_{2})}\;\; {{-g_{\mu \nu}}\over
{q^{2}+i\epsilon}} -\frac {1}{{q^{+}}^{2}} {\bar u}_{e}
(p'_{1})\gamma^{+}{u_{e }(p_{1})}\; {\bar u}_{{\mu}}(p'_{2})\gamma^{+}
{u_{{\mu }}(p_{2})}\right].
\end{equation}
The second term here
 originates from the noncovariant terms in the gauge
propagator.  It is easily shown to be compensated by the
instantaneous contribution
to the matrix element deriving from the corresponding last term in (25),
of abelian QED theory.  The familiar covariant expression for the matrix
element
is then recovered.

\subsection {The $\beta$-Function in Yang-Mills Theory}

In this section we will illustrate the
renormalization procedure in LF-quantized l.c. gauge QCD by an
explicit computation to one-loop,
 for simplicity,  in the
pure nonabelian Yang-Mills theory.
Gross and
Wilczek \cite{gross} and Politzer \cite{polit} computed the
$\beta$-function in QCD from the gluonic vertex in the conventional
theory.  The corresponding LF computation becomes simpler because
the gauge propagator in l.c. gauge is transverse with respect to both
$k^{\mu}$ and the gauge direction $n^{\mu}$, and ghost fields
are absent. \\

\noindent
{\bf Gluon Self-energy corrections:}

\vspace{.5cm}
\begin{figure}[htb]
\begin{center}
\leavevmode
\epsfbox{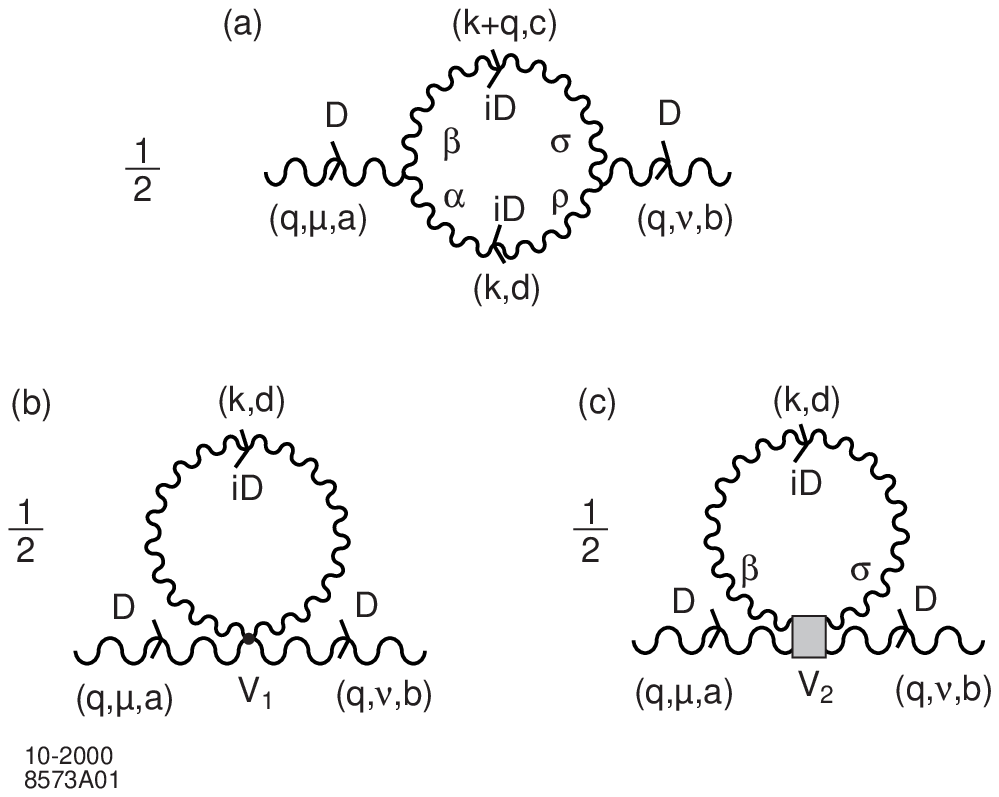}
\end{center}
\caption[*]{Yang-Mills self-energy to one loop.  (a) Gluon self-energy diagram
$\Pi^{\mu\nu}_{ab}(q)$; (b) tadpole diagram containing vertex $V_{1}$,
vanishing in dimensional regularization; (c) non-vanishing
tadpole diagram containing vertex $V_{2}$.}
\label{fig1}
\end{figure}

The propagator modification is given by
\begin{equation}
\delta_{ab}\frac {i D_{\lambda\delta}(q)}{q^{2}} + \delta_{aa'}\frac { i
D_{\lambda\mu}(q)}{q^{2}} \, \Pi^{\mu\nu}_{a'b'}(q) \, \delta_{b'b} \frac
{ i D_{\nu\delta}(q)}{q^{2}} +\cdots
\end{equation}
The
contribution to the gluon polarization tensor $\Pi^{\mu\nu}_{ab}(q)$
coming solely from the three-gluon interaction is (Fig.  \ref{fig1}a)
\begin{eqnarray}
\Pi^{\mu\nu}_{ab}(q)\!&=& \!  \int
\!\frac{d^{d}k}{(2\pi)^{d}} \,\frac{1}{2}\,(-g_{d})^{2} \,f_{adc}\,
f_{bcd} \, F^{\mu\alpha\beta}(-q,-k,k+q)\nonumber\\ &&
\qquad\qquad\qquad F^{\nu\sigma\rho}(q,-k-q,k) \, \frac
{iD_{\beta\sigma}(k+q)}{(k+q)^{2}} \, \frac {iD_{\alpha \rho}(k)}{k^{2}}
\nonumber \\ &=& \frac{1}{2}\, g^{2} \,\delta_{ab}\,C_{A} \;
\Pi^{\mu\nu}(q).
\end{eqnarray}
Here associated to the three outgoing
momenta, $p_{1}^{\lambda}$, $p_{2}^{\mu}$, $p_{3}^{\nu}$, satisfying
$(p_{1}+p_{2}+p_{3})^{\mu}=0$, we define
\begin{equation}
F_{\lambda\mu\nu}(p_{1},p_{2},p_{3})= (p_{1}-p_{2})_{\nu}g_{\lambda\mu}
+ (p_{2}-p_{3})_{\lambda}g_{\mu\nu}+(p_{3}-p_{1})_{\mu}g_{\nu\lambda}
=-F_{\lambda\nu\mu}(p_{1},p_{3},p_{2}),
\end{equation}
use $\,
f_{adc}f_{bcd}= - C_{A} \delta_{ab} $ and write
\begin{equation}
\Pi^{\mu\nu}(q)=\mu^{4-d}\int \frac {d^{d}k}{(2\pi)^{d}}
\frac{1}{[k^{2}+i\epsilon]\, [(k+q)^{2}+i\epsilon]} \; I^{\mu\nu}(q,k),
\end{equation}
with
\begin{eqnarray} I^{\mu\nu}(q,k)&=
&[-(2k+q)^{\mu}g^{\alpha\beta}+(k-q)^{\beta} g^{\mu\alpha}
+(2q+k)^{\alpha} g^{\mu\beta}] D_{\alpha\rho}(k) \nonumber \\
&&[-(2k+q)^{\nu}g^{\rho\sigma}+(k-q)^{\sigma} g^{\nu\rho} +(2q+k)^{\rho}
g^{\nu\sigma}] D_{\sigma\beta}(k+q).
\end{eqnarray}
The dimensionless
coupling is indicated by $g$ while $g_{d}= (\mu)^{\frac{4-d}{2}}g$ and
$\mu$ indicates the mass parameter associated with the dimensional
regularization which we will be adopting.

We first note that every internal gluon line
carries a factor $ D_{\mu\nu}$, and the polarization vector
of an external gluon is $E^{\mu}_{(\perp)}
= -D^{\mu}_{\perp}$.  The object of
interest relevant in the renormalization of the theory under
consideration is clearly the combination $D_{\lambda\mu}(q) \,
\Pi^{\mu\nu}_{ab}(q) \, D_{\nu\delta}(q)$.  We may therefore use the
transversity properties of $D^{\mu\nu}(q)$ to simplify the original
expression and consider instead the following reduced expression for
$I^{\mu\nu}$ in the integrand
\begin{eqnarray}
I^{\mu\nu}(q,k)&=&
[-(2k+q)^{\mu}g^{\alpha\beta}-2q^{\beta} g^{\mu\alpha} +2q^{\alpha}
g^{\mu\beta}] D_{\alpha\rho}(k) \nonumber \\
&&[-(2k+q)^{\nu}g^{\rho\sigma}-2q^{\sigma} g^{\nu\rho} +2q^{\rho}
g^{\nu\sigma}] D_{\sigma\beta}(k+q).  \end{eqnarray} Explicitly
\begin{eqnarray} I^{++}&=& 2(2k+q)^{+}(2k+q)^{+} \nonumber \\
I^{+\perp}&=& I^{\perp+}= 2(2k+q)^{+}(2k+q)^{\perp }+ 2(2k+q)^{+}
\left[\frac{1}{(k+q)^{+}}-\frac{1}{k^{+}}\right]S_{\perp} \nonumber \\
I^{\perp\perp'}&=& 2\,(2k+q)^{\perp}(2k+q)^{\perp'} \nonumber \\ &&-
2\,[(2k+q)_{\perp} S_{\perp'}+(2k+q)_{\perp'} S_{\perp}]\;
\left[\frac{1}{(k+q)^{+}}-\frac{1}{k^{+}}\right] \nonumber \\ &&- 4
\,g^{\perp\perp'}\, S_{\perp}S_{\perp} \left[\frac{1}{{(k+q)^{+}}^{2}}+
\frac{1}{{k^{+}}^{2}}\right] \nonumber \\ &&- 8 \,S_{\perp} S_{\perp'}
\,\frac{1}{k^{+}\,(k+q)^{+}}
\end{eqnarray}
where
\begin{eqnarray}
S_{\mu}\equiv S_{\mu}(k,q)&=& (k_{\mu}q^{+}-q_{\mu}k^{+}) \nonumber \\
&=&-S_{\mu}(q,k)= S_{\mu}(k+a q, q)=S_{\mu}(k, q+b k)
\end{eqnarray}
Here the properties $D^{\perp}_{\rho}(k)=
[-\delta^{\perp}_{\rho}+(k^{\perp}/k^{+})\delta^{+}_{\rho}]$,
$\,D^{\mu\nu}(k)D_{\mu\nu}(k)=2$, $\,D_{\lambda\mu}(q){D^{\mu}}_{\perp}
(k)= -D_{\lambda\perp}(q)$ were used to simplify the expressions.

In order to carry out the renormalization procedure, we will need to
isolate the divergent terms in the matrix element.
We will adopt
dimensional regularization since it preserves all gauge symmetries.
The singularities in the Feynman
propagators of the
dynamical components $A_{\perp}$ and $\psi_{+}$ are given
by the standard causal prescription.
The $1/k^{+}$ singularity will be handled by the
Mandelstam \cite{mand} and Leibbrandt \cite{leibb1} prescription in the
l.c. gauge.  A derivation of this prescription
has also been given \cite{Bassetto:1997ba} in the context of
equal-time canonical quantization.  One can also justify the
Mandelstam-Leibbrandt (ML) procedure by noting that in a two-dimensional
massless theory on the LF,
the causal prescription for the $k^{2}\approx 0$ singularity in
$1/k^{2}\equiv 1/(2k^{+}k^{-})$ is identical to that
given by the causal ML prescription for the $1/k^{+}$ singularity.  Since
we wish to have consistent analytic continuation in the number of
dimensions $d$, the dimensional
regularization plus ML prescription appears be a mathematically sound
procedure.

The ML prescription is often written as
\begin{equation}
\frac{1}{q\cdot n}= \lim_{\epsilon\to 0}\frac{(q\cdot
n^{*})}{(q\cdot n)(q\cdot n^{*}) +i\epsilon}\,
\end{equation}
where
$\epsilon\to 0+ \,$ and the light-like four-vector $n^{*}_{\mu} $
represents the dual of $n_{\mu}$ with the components given by,
$n^{*}_{\mu}=\delta^{-}_{\mu}$.  We recall that such a pair of null
vectors, $n_{\mu}$ and $n^{*}_{\mu} $, arise quite naturally in the LF
framework, for example, in the LF quantized QCD in covariant gauge
\cite{cov}, when we define\footnote{\baselineskip 12pt
$e^{(+)}=(1, {\vec k}/k^{0})/{\sqrt
2}$, $e^{(-)}=(1, -{\vec k}/k^{0})/{\sqrt 2}$, $e^{(1)}=(0, {\vec
\epsilon}(k; 1))$, $e^{(2)}=(0, {\vec \epsilon}(k; 2))$ where
${\vec\epsilon}(k;1)$, ${\vec\epsilon}(k;2)$ and ${\vec k}/|{\vec k}|$
constitute the usual orthonormal set of three-vectors.  In l.c. gauge
$\, n\cdot n^{*}=1$, $n\cdot n=n^{*}\cdot n^{*}=0$ and associated to any
four-vector $q_{\mu}$ we may define the four vectors $\,q_{||}$ and
$\,q_{(\perp)}$ by $\,{q_{||}}_{\mu} \,(n\cdot n^{*})=(n^{*}\cdot q \,
n_{\mu}+ n\cdot q \,n^{*}_{\mu})$ and $\,{q_{(\perp)}}_{\mu}=
q_{\mu}-{q_{||}}_{\mu}$.  } the linearly independent set of four gauge
field polarization vectors.

Unlike the principal value
prescription for $1/k^{+}$, which would enter in conflict with the
causal prescription for $1/k^{2}$,
the causal $n^{*}_{\mu}$- prescription is consistent with both
Wick rotation and power counting \cite{leibb, Bassetto:1997ba}.

The divergent part of $\Pi^{\mu\nu}$ may be computed straightforwardly
employing the available list of integrals \cite{leibb, Bassetto:1997ba}.
We find
\begin{eqnarray}
div\;\Pi^{++}&=& \frac{2}{3} \,{q^{+}}^{2}\,
I^{div} \nonumber \\ div \; \Pi^{+\perp}&=& -\frac{10}{3}\, q^{+}
q^{\perp}\, I^{div} \nonumber \\ div \; \Pi^{\perp\perp'}&=&
2\,\left[\frac{11}{3} (q^{2} g^{\perp\perp'}- q^{\perp}q^{\perp'})- 8\,
q^{+}q^{-} g^{\perp\perp'} \right]\, I^{div}
\end{eqnarray}
Here
$\,(2\pi)^{4} I^{div}= 2i\pi^{2}/{(4-d)}\to i\pi^{2} (2/\epsilon)$, with
$d=(4-\epsilon)$, $\epsilon\to 0+$,  is the pole term in the divergent
integral
\begin{equation}
(2\pi)^{4}\; \mu^{4-d} \int \frac
{d^{d}k}{(2\pi)^{d}} \;\frac{1} {k^{2}(k-q)^{2}}= i \pi^{2} \;\left[
N_{\epsilon}- \ln \frac {-q^{2}}{\mu^{2}} +\cdots
\right] + o(\epsilon)
\end{equation}
where

$$ N_{\epsilon}= \left[
\frac{2}{\epsilon}- \gamma_{E}+ \ln (4\pi)\right] $$

\noindent The expressions of $\, \Pi^{++}$, $\Pi^{+\perp}$ agree with the
corresponding expressions computed in Ref.  \cite{leibb}, in the
conventional l.c. gauge QCD, where a different gluon propagator is used.
However, the expression for $\Pi^{\perp\perp'}$ is found to be
different.  As a consequence in the LF quantized
theory we find covariant as well as noncovariant divergent terms in
\begin{equation}
D_{\lambda\mu}(q) \,
\Pi^{\mu\nu}_{ab}(q)\, D_{\nu\delta}(q) = \frac{ g^{2}}{16 \pi^{2} }
C_{A}\delta_{ab}\, (-\frac{11}{3} q^{2} + 8 q^{+}q^{-})
\,\;i\, \left[
N_{\epsilon}- \ln \frac {-q^{2}}{\mu^{2}} +\cdots
\right] \; D_{\lambda\delta}(q)
\end{equation}
Also, in view of the properties of $D_{\mu\nu}$, the
computation of (39) does not require the evaluation of components
other than those given
in (37).  We note also that
\begin{equation}
q_{\nu} I^{\mu\nu}= 2\,
(q^{2}+2k.q)\,\, \left[ (2k+q)^{\mu} + (q^{\beta}g^{\mu\alpha}-
q^{\alpha} g^{\mu\beta})\right]\,\,{D^{\rho}}_{\beta}(k+q)
D_{\rho\alpha}(k).
\end{equation}
The corresponding divergent part is shown to be
\begin{equation}
div\; q_{\nu}\ \Pi^{\nu\mu}(q)= -8\, q^{-}
{q^{+}}^{2}\, {D^{\mu}}_{+}(q)\;I^{div},
\end{equation}
which allows us
to compute $ div\,\, \Pi^{-+}(q)= \,(\frac{10}{3} q^{2}- \frac{22}{3}
q^{+}q^{-}) \,\, I^{div}$ by setting $\mu=+$.

The result (39) obtained here is different from the earlier l.c. gauge
computations \cite{leibb} in the conventional framework.  The
noncovariant piece $\, 8 q^{+}q^{-} \, N_{\epsilon}$,  in (39),
however, is compensated
by an equal contribution but with opposite sign
which arises from the {\it tadpole} graphs.
The connection between tree-graph computations in the conventional
light-cone gauge formalism and light-cone time-ordered perturbation
theory have been discussed in
refs.
\cite{Paston:1999fq,Harindranath:1993de,Zhang:1993dd,Ligterink:1995tm}

The computation of the loop corrections to the gluon
self-energy illustrates the essential difference between the light-front
and conventional formalisms.  There are two {\it tadpole} graphs
(Figs.
\ref{fig1}b and
\ref{fig1}c) to be considered.
The one associated with the four-gluon coupling $V_{1}$ gives a vanishing
result, but the contribution coming from the instantaneous interaction
$V_{2}(p_{1},p_{2},p_{3},p_{4})$ is found to be nonvanishing in
dimensional regularization due to the momentum dependence of the vertex
itself.
The divergent part of the matrix element is easily reduced to
\begin{eqnarray}
&&\mu^{4-d}\int \frac
{d^{d}k}{(2\pi)^{d}}\,\frac{ 1}{2} (ig^{2}) C_{A} \delta_{ab}
\left[-(\frac {k^{+}-q^{+}}{k^{+}+q^{+}})^{2} - (\frac
{k^{+}+q^{+}}{k^{+}-q^{+}})^{2}\right] \frac{iD^{\mu \nu}(k)}{k^{2}}
D_{\lambda\mu}(q) D_{\nu\delta}(q) \nonumber \\ &=& \frac{ g^{2}}{16
\pi^{2} } \,C_{A}\,\delta_{ab}\, (- 8 q^{+}q^{-})
\,\;i\, \left[N_{\epsilon}- \ln \frac {-2 q^{+}q^{-}}{\mu^{2}} +\cdots
\right]
\;\, D_{\lambda\delta}(q)
\end{eqnarray}
Here we made use of the
useful identity $ D_{\lambda\mu}(q) \,D^{\mu\nu}(k)\,D_{\nu\delta}(q) =
D_{\nu\delta}(q)$ to arrive at the second line.  The usual shift
operation in dimensional regularization
is used to bring the integral to another type of basic
divergent integral \cite{leibb} which is, however,  noncovariant.
On adding the tadpole contribution (42)
to (39) the net
coefficient of $ (N_{\epsilon}+ \ln \mu^{2} )$
in the gluon self-energy correction is covariant, since
the noncovariant $q^{+}q^{-}$ terms mutually cancel (see also Appendix
C).
Retaining only the pole term we find
\begin{equation}
D_{\lambda\mu}(q) \, \Pi^{\mu\nu}_{ab}(q)\,
D_{\nu\delta}(q) = \frac{ g^{2}}{16 \pi^{2} } C_{A}\delta_{ab}\,
(-\frac{11}{3} q^{2} ) \,\;i\,
(\frac{2}{\epsilon})\, \,D_{\lambda\delta}(q).
\end{equation}
The divergent part in (43) is $\propto q^{2}$ which ensures that
the vanishing gluon mass
 remains unaltered due to the one-loop gluon self-energy
correction.

The multiplicative renormalization constant $Z_{3}$ which corrects
the gluon propagator is defined by
\begin{equation}
Z_{3}
\,\;\delta_{ab}\frac {i D_{\lambda\delta}(q)}{q^{2}}= \delta_{ab}\frac
{iD_{\lambda\delta}(q)}{q^{2}} + \delta_{aa'} \frac { i
D_{\lambda\mu}(q)}{q^{2}} \, \Pi^{\mu\nu}_{a'b'}(q) \, \delta_{b'b} \frac {
i D_{\nu\delta}(q)}{q^{2}} +\cdots
\end{equation}
 and we obtain
\begin{equation} Z_{3}= 1+\frac{ g^{2}}{16 \pi^{2} }\, C_{A}
\,\frac{11}{3} \, \,
(\frac{2}{\epsilon})
\end{equation} \\

\nl {\bf Vertex corrections:}

\vspace{.5cm}
\begin{figure}[htbp]
\begin{center}
\leavevmode
 \epsfbox{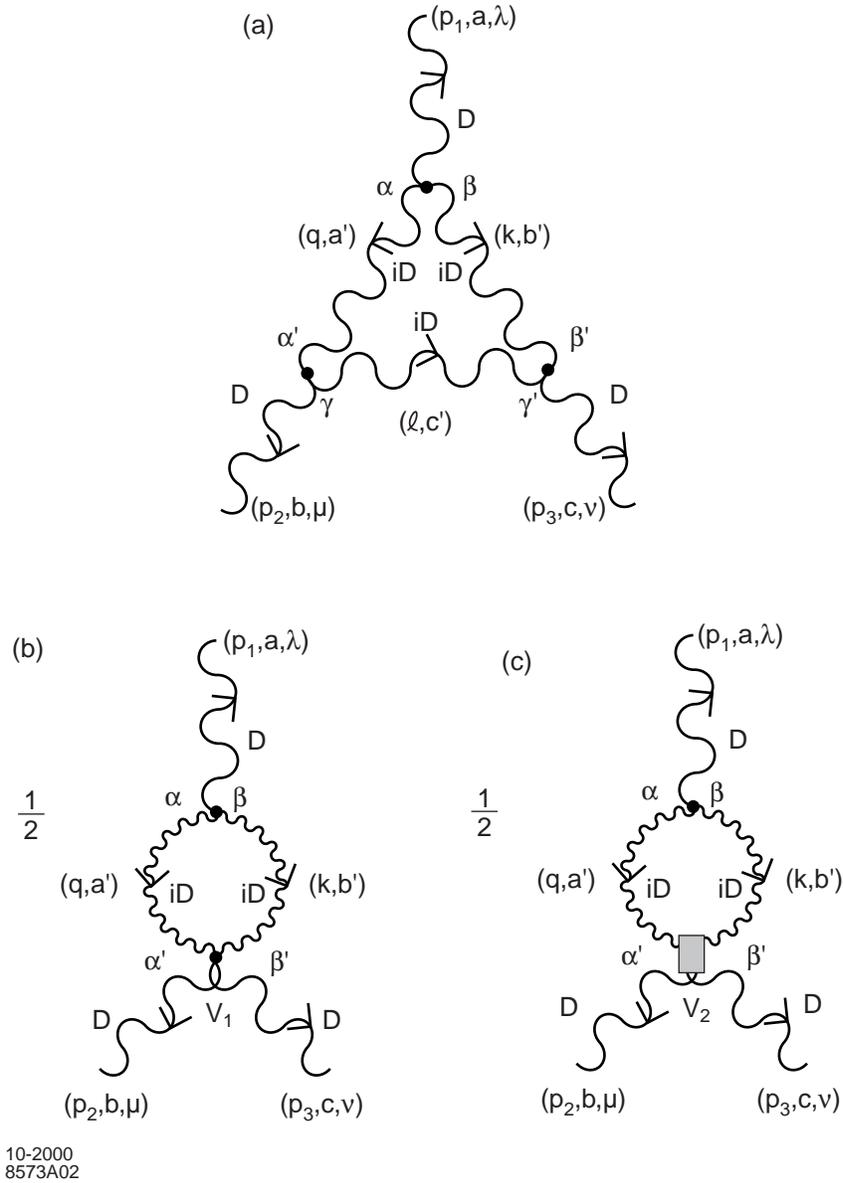}
\end{center}
\caption[*]{
Three-gluon vertex diagrams. (a) Triangle diagram; (b) swordfish
diagram containing vertex $V_{1}$; the other two diagrams are
obtained by cyclic permutations of the external line indices; (c) swordfish
diagrams containing vertex $V_{2}$;  the other two diagrams are
obtained by cyclic permutations of the external line indices.
}
\label{fig2}
\end{figure}

In pure Yang-Mills theory the gluon vertex corrections to one-loop arise
from the three-gluon interaction alone, the triangle diagram,
 (Fig. \ref{fig2}a) and from the two types of
{\it
swordfish} graphs (Figs. \ref{fig2}b and \ref{fig2}c) in which one of the
two vertices
carries a four-gluon interaction, which may be of type $V_{1}$ or
type $V_{2}$,
 while the other one contains a three-gluon interaction.
The complete vertex to order $g^{3}$ is written as
\begin{equation}
-g
f_{abc} {\cal F}_{\lambda\mu\nu} = -g f_{abc} \,[ F_{\lambda\mu\nu}+
\Delta_{\lambda\mu\nu}](p_{1},p_{2},p_{3})= -g f_{abc }
\,F_{\lambda\mu\nu}(p_{1},p_{2},p_{3})\, (1+ \bar\Delta)  \ .
\end{equation}

Consider first the triangle diagram.
Since, as remarked before, each gluon line carries with it a factor of
$D$ we will simplify the expressions right from the start
making use also of the presence of the factor
$\,D_{\lambda'\lambda}(p_{1})D_{\mu'\mu}(p_{2}) D_{\nu'\nu}(p_{3}) $,
coming from the external gluon lines.
 The matrix
element for the one loop correction to order $g^{2}$ is written as
\begin{equation}
(-g_{d}^{2})\,(-\frac{1}{2}C_{A}\,
f^{abc}) \,(i)^{3} \;T^{\lambda\mu\nu}(p_{1},p_{2},p_{3})
\end{equation}
where we have use $f_{aa'b'}f_{bb'c'}f_{cc'a'}= (C_{A}/2)f_{abc}$.  A factor
of $i$ comes from each of the gluon propagators and the expression for
$T^{\lambda\mu\nu}(p_{1},p_{2},p_{3})$ is given by (Fig. \ref{fig2}a)
\begin{eqnarray}
\int \frac{d^{d}
q}{(2\pi)^{d}}&&\frac{D_{\alpha\alpha'}(q)
D_{\beta\beta'}(k)D_{\gamma\gamma'}(l)}{(q^{2}+i\epsilon)(k^{2}+i\epsilon)
(l^{2}+i\epsilon)}
[(p_{1}-q)^{\beta}g^{\alpha\lambda}+(q-k)^{\lambda}g^{\alpha\beta}
+(k-p_{1})^{\alpha}g^{\lambda\beta}]\nonumber\\
&&\qquad\qquad[(-q-p_{2})^{\gamma}
g^{\alpha'\mu}+(p_{2}-l)^{\alpha'}g^{\gamma\mu}
+(l+q)^{\mu}g^{\alpha'\gamma}]\nonumber \\
&&\qquad\qquad[(-k+l)^{\nu}g^{\beta'\gamma'}+
(-l-p_{3})^{\beta'}g^{\nu\gamma'} +(p_{3}+k)^{\gamma'}g^{\beta'\nu}]
\end{eqnarray}
where $ k=-(q+p_{1})$, $ l=(q-p_{2})$, $
p_{1}+p_{2}+p_{3}=0$ and the $D's $ associated with the external gluon lines
are understood.

Proceeding as before we may consider instead the following reduced
expression
\begin{eqnarray}
 8 \int \frac{d^{d}
q}{(2\pi)^{d}}&&\frac{D_{\alpha\alpha'}(q)
D_{\beta\beta'}(k)D_{\gamma\gamma'}(l)}{(q^{2}+i\epsilon)(k^{2}+i\epsilon)
(l^{2}+i\epsilon)} [p_{1}^{\beta}g^{\alpha\lambda} +
q^{\lambda}g^{\alpha\beta} -p_{1}^{\alpha}g^{\lambda\beta}] \nonumber\\
&&[-p_{2}^{\gamma}g^{\alpha'\mu}+p_{2}^{\alpha'}g^{\gamma\mu} +
q^{\mu}g^{\alpha'\gamma}]
[-k^{\nu}g^{\beta'\gamma'}-p_{3}^{\beta'}g^{\nu\gamma'}
+p_{3}^{\gamma'}g^{\beta'\nu}]
\end{eqnarray}
The divergent terms in
$T^{\lambda\mu\nu}$ are then easily identified and may be rewritten as
follows
\begin{eqnarray}
8 \int \frac{d^{d}
q}{(2\pi)^{d}}\frac{1}{q^{2}k^{2}l^{2}}
&&\left[2q^{\lambda}q^{\mu}k^{\nu}+p_{3}^{\alpha}q^{\lambda}q^{\mu}
\{D^{\nu}_{\alpha}(l)-D^{\nu}_{\alpha}(k) \}\right.  \nonumber\\
&&\left.  -k^{\nu}q^{\lambda}p_{2}^{\alpha}\{D^{\mu}_{\alpha}(q)-
D^{\mu}_{\alpha}(l)\}
-k^{\nu}q^{\mu}p_{1}^{\alpha}\{D^{\lambda}_{\alpha}(k)-
D^{\lambda}_{\alpha}(q)\} \right].
\end{eqnarray}
The contribution to
the divergent part comes solely from the first term, since the divergent
terms coming from the noncovariant terms vanish.
The divergent part of the one-loop correction from the triangle diagram
has the form
$\,-g f_{abc }\,F_{\lambda\mu\nu}(p_{1},p_{2},p_{3})\,
 {\bar\Delta}_{1}$ with
\begin{equation}
\bar\Delta_{1} = \frac{ g^{2}} {16 \pi^{2}} C_{A}
\;(\frac{1 }{3})\;\left[
N_{\epsilon}- \ln \frac {Q^{2}}{\mu^{2}} +\cdots
\right]
\end{equation}
where $\,Q^{2}=-(p_{2}+p_{3})^{2}$.
It is covariant and different from the one
found \cite{Bassetto:1997ba,leibb} in equal-time l.c. gauge framework
where a different propagator is used, and the corresponding expressions
contain noncovariant pieces.

The total contribution from each type of {\it swordfish} diagram
comes from three similar diagrams, the two others being
constructed from the first one by cyclic
permutations of the set of labels of the three external gluon lines.
 The net divergent contribution
following from the two types of
diagrams is given by $\,-g f_{abc }\,F_{\lambda\mu\nu}(p_{1},p_{2},p_{3})
\,  {\bar\Delta}_{2}$ with
\begin{equation}
\bar\Delta_{2} = \frac{
g^{2}} {16 \pi^{2}} C_{A} \; (-4) \; \left[
N_{\epsilon}- \ln \frac {Q^{2}}{\mu^{2}} +\cdots \right]
\end{equation}
 The noncovariant terms cancel out leading to the
covariant result ${\bar\Delta}_{2}$ (see below).
 The gluon
vertex renormalization constant $Z_{1}$ is defined by, $\bar\Delta=
(\bar\Delta_{1}+\bar\Delta_{2})$,
\begin{equation}
-g f_{abc} {
F}_{\lambda\mu\nu}(p_{1},p_{2},p_{3})\;\,\frac{1}{Z_{1}} = -g
f_{abc} { F}_{\lambda\mu\nu}(p_{1},p_{2},p_{3})\;\,(1\,+\,\bar\Delta)
\end{equation}
and we obtain 
\begin{equation}
\frac{1}{Z_{1}}=
1-\frac{g^{2}}{16\pi^{2}}\, C_{A} \,(\frac{11}{3} )\,
(\frac{2}{\epsilon})
\end{equation}

 We find $Z_{1}=Z_{3}$ in our doubly transverse gauge framework
in the l.c. gauge LF quantized theory.
 The gauge coupling constant renormalization constant $Z_{g}$,
is defined by $ \,Z_{g}= {Z_{1}}/{(Z_{3})^{3/2}}$.  In the lowest order
perturbation theory it has the form
\begin{equation}
Z_{g}=\frac {Z_{1}}{(Z_{3})^{3/2}}=(Z_{3})^{-\frac{1}{2}}
\approx \, 1-\frac{g^{2}}{16\pi^{2}} \,C_{A} \,(\frac{11}{6})\; \, \,
(\frac {2}{\epsilon}) = 1- {g^{2}} \, \beta_{0}\,
(\frac{1}{\epsilon})
\end{equation}
where $\,\beta_{0}= (1/{16\pi^{2}})\; ({11 \,C_{A}}/{3})\, >0$.
It agrees with the result found \cite{gross,
polit} in QCD, in the conventional {\it instant form} framework
 when the quark fields are ignored \cite{ref:gg}.
The computation of $\beta$-function and the discussion of the
asymptotic freedom is made following the standard procedure \cite {collins,
lang}.

A sketch of the computation involved in the {\it swordfish } diagrams
is presented here.  The matrix element simplifies considerably due to
the nice properties of the $D's$ leading to the reduced form
\begin{eqnarray}
&& \frac{1}{2}\,(-ig_{d}^{2})\, \frac{1}{2} C_{A}\,
f_{abc}\, \int \frac{d^{d} q}{(2\pi)^{d}}\frac{1}{q^{2}k^{2}}
iD_{\alpha\alpha'}(q)\, iD_{\beta,\beta'}(k) \nonumber \\ &&
2\left[p_{1}^{\beta}g^{\alpha\lambda}-p_{1}^{\alpha}g^{\beta\lambda}
+q^{\lambda} g^{\alpha\beta}\right]\, \left[A g^{\alpha'\mu}
g^{\beta'\nu} +B g^{\alpha'\nu} g^{\beta'\mu}+ C g^{\alpha'\beta'}
g^{\mu\nu}\right]
\end{eqnarray}
where
\begin{equation}
A= 3-
\frac{(q+p_{2})^{+}(p_{3}+k)^{+}} {(q-p_{2})^{+\,2}}, \quad B= -3
+\frac{(k+p_{2})^{+}(p_{3}+q)^{+}} {(q-p_{3})^{+\,2}}, \quad C= 2
\frac{(k-q)^{+}(p_{3}-p_{2})^{+}} {(q+k)^{+\,2}}
\end{equation}
with a
multiplication by the factor $D_{\lambda'\lambda}(p_{1})
D_{\mu'\mu}(p_{2})D_{\nu'\nu}(p_{3}) $ being understood.  The integrand
may be simplified further and recast as
\begin{eqnarray}
&& p^{\rho}
\left[\, -g^{\widetilde\lambda\widetilde\mu}\, (\,A
D_{\rho}^{\widetilde\nu}(k)- B
D_{\rho}^{\widetilde\nu}(q)\,)\, +g^{\widetilde\lambda\widetilde\nu} \, (\,A
D_{\rho}^{\widetilde\mu}(q)- B D_{\rho}^{\widetilde\mu}(k)\,) \right.
\nonumber
\\ && \left.  + C g^{\mu\nu}\, (\,A D_{\rho}^{\widetilde\lambda}(q)-
D_{\rho}^{\widetilde\lambda}(k) \,)\right]+ \,q^{\lambda}
g^{\widetilde\mu\widetilde\nu}\, (A+B)\,+\,2 C\, q^{\lambda}\,g^{\mu\nu}
\end{eqnarray}
where $\widetilde\mu \equiv {\perp} = 1,2 $ only.  The
divergent terms then can be picked up straightforwardly.  Some of them
drop out if we use also $ p_{1}^{\lambda} D_{\lambda'\lambda}(p_{1})=0$
and $D_{-\mu}=0$.
When we add to this result the two
expressions obtained from the present one by the cyclic
permutations of 3-tuples $(p_{1},\lambda, a)$, $(p_{2},\mu, b)$, and
$(p_{3},\nu, c)$, the noncovariant pieces drop out, leading to the
covariant expression for $\bar\Delta_{2}$ given above.

The divergent terms arising from each type of the {\it swordfish} graphs
contain covariant as well as noncovariant pieces.
The contribution corresponding to
the usual four-gluon vertex $V_{1}$ in our framework agrees with
the one found in the earlier l.c.  gauge computations
\cite{Bassetto:1997ba, leibb}, in the conventional framework.  It thus
gives us a consistency check of our calculations, since the computation
here is insensitive to the $n_{\mu}n_{\nu}k^{2}/{k^{+}}^{2}$ term in our
gauge propagator.
The doubly transverse gauge propagator of the LF quantized theory
greatly simplifies the computation: in compensation for the few extra
interaction vertices,  there are no ghost interactions. \\

\subsection {Gluon Self-Energy corrections from Quark loop}

For each flavor $f$ of quark the net gluon self-energy contribution
arising from a quark loop is found to acquire the usual form
\begin{equation}
\Pi^{(F)\mu\nu}_{aa'}= (-1)\, (ig)^{2}\,
\sum_{ij} t^{a}_{ij} t^{a'}_{ji}\,\,
\mu^{4-d}\, (i)^{2} \,\int \frac {d^{d}k}{(2\pi)^{d}} \;
\frac {Tr({\not k}+m_{f})\gamma^{\mu} ({\not k}+{\not
q}+m_{f})\gamma^{\nu}}{[k^{2}-{m_{f}}^{2}]\, [(k+q)^{2}-{m_{f}}^{2}]}.
\end{equation}
where the factor $D_{\lambda\mu}(q)
D_{\nu\delta}(q)$ is understood.
It is easily shown that
the noncovariant contributions either vanish or
are mutually cancelled.  Hence
\begin{eqnarray}
 div
\;D_{\lambda\mu}(q)\,\Pi^{(F)\mu\nu}_{aa'}\, D_{\nu\delta}(q)&= &
 \frac {g^{2}}{16 \pi^{2}}\,\, \delta_{aa'}\,\, T_{F}\,\,
(\,\frac{4}{3}\, q^{2} )\; D_{\lambda\delta}(q)
\nonumber \\
&& i\, \left[ N_{\epsilon}- 6\int_0^1 dx \, x(1-x)\, \ln \,\frac
{m_{f}^{2} -x(1-x)\, q^{2}}{\mu^{2}} \right]
\end{eqnarray}
where $\sum_{ij} t^{a}_{ij} t^{a'}_{ji}= T_{F}\delta_{aa'}$.
The contributions from
$\Pi^{(F)-\nu}_{aa'}$ or $\Pi^{(F)\mu -}_{aa'}$ are automatically suppressed
in view of $D_{\mu -}= 0$, as they should, since $A_{-}=0$ in the l.c.
gauge.  We have used also
\begin{equation}
D_{\lambda\mu}(q)\, \,(q^{2}
g^{\mu\nu}-q^{\mu}q^{\nu})\, \,D_{\nu\delta}(q) = - q^{2}
D_{\lambda\delta}(q).
\end{equation}
For the total number $n_{f}$ of massless quarks we find
\begin{equation}
 div
\;D_{\lambda\mu}(q)\,\Pi^{(F)\mu\nu}_{aa'}\, D_{\nu\delta}(q)=
 \frac {g^{2}}{16 \pi^{2}}\, \delta_{aa'}\, T_{F}\,
(\frac{4}{3} \, n_{f}\,q^{2} )
\, i\, \left[ N_{\epsilon}- \ln \frac{-q^{2}}{\mu^{2}}
+ \cdots \right]\; D_{\lambda\delta}(q).
\end{equation}
 The net contribution to coefficient of the pole term in the gluon self-energy
becomes
\begin{eqnarray}
 div \; D_{\lambda\mu}(q)\,
\left[\Pi^{\mu\nu}_{aa'}+\Pi^{(F)\mu\nu}_{aa'}\right]
\, D_{\nu\delta}(q)&=&
 \frac {g^{2}}{16 \pi^{2}}\, \delta_{aa'}\,\, q^{2} \,\,
\left( -\frac{11}{3} C_{A} +
\frac{4}{3} \, n_{f}\, T_{f}\right) \nonumber \\
 && \quad i\, (\frac {2}{\epsilon})
\, D_{\lambda\delta}(q).
\end{eqnarray}
which leads to
\begin{equation}
Z_3 =1+ \frac {g^{2}}{16 \pi^{2}}\, \left( \frac{11}{3} C_{A} -
\frac{4}{3} \, n_{f}\, T_{F}\right) \, (\frac{2}{\epsilon})
\end{equation}
with
\begin{equation}
{Z_3 }^{-\frac{1}{2}}=1- \frac {g^{2}}{16 \pi^{2}}\,
\left( \frac{11}{6} C_{A} -
\frac{2}{3} \, n_{f}\, T_{F}\right) \,
(\frac{2}{\epsilon})
\end{equation}
in our ghost free l.c. gauge LF quantized QCD framework.
We also
did not perform any renormalization of the gauge parameter,
required in the conventional covariant gauge theory framework.
A complete discussion of QCD in our framework
will be given elsewhere \cite{ref:dd}.

\section{Conclusions}

The canonical quantization of l.c. gauge QCD in the {\it front form}
theory has been derived employing the Dirac procedure to construct a
self-consistent LF Hamiltonian theory.  The formulation begins with the
gauge-fixed BRS invariant quantum action, but the final result is
ghost-free.
The interaction
Hamiltonian is obtained in a simple form by retaining the dependent
components $A_{+}$ and $\psi_{-}$.
Its form closely resembles the
interaction Hamiltonian of covariant theory, except for the presence of
instantaneous interactions which are analogous to the Coulomb interactions
in transverse gauge.
The Dyson-Wick perturbation theory expansion based on
equal-LF-time ordering is then constructed in a manner which
allows one to perform high order computations in straightforward
fashion.  Our formulation of the light-front Hamiltonian matches
the rules of light-front-time-ordered perturbation
theory \cite{Lepage:1980fj}.

In our formulation of gauge theory, the free gauge field satisfies
the Lorentz condition as an operator equation as well as the
light-cone gauge constraint.  Since the propagator of the massless gauge
field is doubly transverse both with respect to the
four-momentum and the four-vector $n_{\mu}$ of gauge direction,
the propagating gluons have only two
physical degrees of freedom, and no ghost fields need to be considered.
The physical transverse polarization vectors of the gluons may be
conveniently identified as $E^{\mu}_{(\perp)}(k)\equiv
-D^{\mu}_{\perp}(k)$ so that each gluon line,
external or internal, carries with it a factor of $D_{\mu\nu}$.
The remarkable properties of these factors give rise to much simplification
already at the start of computations.

Unitarity relations \cite{ref:uu}
such as the optical theorem are manifest within
each Feynman diagram, rather than as a consequence of
cancelations over sets of diagrams \cite{Papavassiliou:1997fn}.
This allows one to construct effective charges analogous to the
Gell Mann-Low
running coupling of QED based on the structure of self-energy diagrams
using the pinch technique\cite{papa}.  Since the absorptive part of these
contributions are based on physical cross sections, one can define a
physical and analytic renormalization scheme for QCD.

The instantaneous interactions, in fact, are interesting by themselves.
For
example, the semi-classical limit of Thomson scattering is revealed
at the tree level in the l.c. or covariant gauge \cite{cov} on
the LF.  This is relevant since a systematic procedure to obtain the
semi-classical limit seems to be lacking in the {\it front form}
theory.
The LF framework may be useful also for obtaining
the non-relativistic limit of a relativistic field theory, for example, in
the context of chiral perturbation theory.  This is an alternative to
the conventional framework where a functional integral technique
and the Foldy-Wouthuysen transformation is employed.

It is worth stressing that the doubly transverse
propagator obtained here follows from the straightforward application of
the well-tested standard Dirac method for constructing Hamiltonian
formulation for constrained dynamical systems.  It differs from the
singly transverse propagator usually used in the literature in the
context of equal-time l.c. gauge QCD.  In the equal-time formalism the
last term
$n^\mu n^\nu k^2\over n \cdot k n \cdot k$ of the doubly-transverse gauge
propagator is absent so that both transverse and longitudinal gluonic
modes propagate.  The
instantaneous interaction terms generated by l.c. gauge in the LF are
also not present in the analysis of
{\it instant form} noncovariant gauge QCD based on functional
integral quantization.  Even when one allows for counterterms obtained
by
imposing \cite{Bassetto:1997ba,leibb,schweda}
constraints of covariance or by requiring an extended BRS
symmetry, the gauge propagators employed in that framework do not
possess the very useful properties
carried by the doubly transverse LF framework.  We
have illustrated the correspondence between the two formulations of
light-cone gauge with a simple tree-graph calculation of electron-muon
scattering in QED.  The instantaneous terms in the LF interaction
Hamiltonian restore the manifest Lorentz covariance of the matrix element
which was broken by the noncovariant gauge and the noncovariant terms in
the doubly-transverse gauge propagator.  The same is found true in
nonabelian gauge theory.

The singularities in the noncovariant pieces of the field propagators
may be defined using the causal
l.c. gauge ML prescription for
$1/k^{+}$.  The power-counting rules in l.c. gauge then become similar
\cite{leibb2, Bassetto:1997ba,BRS} to those found in covariant gauge
theory as the illustrations show.
We have 
demonstrated explicitly
the equality $Z_{1}=Z_{3}$ and $Z_{g}= {(Z_{3})}^{-\frac{1}{2}}$
at one loop in Yang-Mills theory.  This is expected to be
true in higher orders as well.
Also because of the
Slavnov-Taylor identities \cite{lang},
the corrections to the $qqg$ vertex would be
compensated by the ones arising from the quark field renormalization;
such that the coupling constant renormalization arises only from
the gluonic self-energy.


Computations in our ghost-free framework
require comparable effort as calculations in covariant gauge because
of the remarkable simplifications
arising due to the doubly transverse l.c. gauge propagator.
Higher-loop
computations should be possible in our formalism
by making advantageous use of the
techniques \cite{leibb2} which have recently been developed to handle
multi-loop integrals involving noncovariant integrands.

The
renormalization procedure of LF quantized gauge theory in l.c.g. is thus
similar to that of conventional covariant gauge theory.
The additional interaction terms are in a sense the
appropriate counterterms
which arise naturally in the
canonical quantization in the LF framework.
It is straightforward to show this in QED; a
complete discussion for QCD will be given elsewhere \cite{ref:dd}.

It is worth remarking that we have made an {\sl ad hoc} choice of only
one (of the family) of the characteristic LF hyperplanes,
$x^{+}=const.$, in order to quantize the theory.  The conclusions
reached here and in the earlier works \cite{pre1, cov} confirm the
conjecture \cite{pre1} concerning the irrelevance in the quantized
theory of the fact that the hyperplanes $x^{\pm}=0$ constitute
characteristic surfaces of hyperbolic partial differential equation.
The Hamiltonian version can clearly be implemented in DLCQ
\cite{Pauli:1985ps} which has been shown \cite{ref:cc} to have a
continuum limit.  There is no loss of causality in DLCQ when the
infinite volume limit is properly handled \cite{hari}.  We also note
that nonperturbative computations are often done on the LF in the closely
related (l.c.) gauge $\partial_{-}A_{-} \approx 0$, such as
 to demonstrate \cite{pre1c} the existence of the {\it condensate }
or $\theta$-vacua in the Schwinger model.

\bigskip

\begin{center}{\Large \bf Appendix A}  \end{center}
\bigskip

\nl {\bf The Feynman Rules}\par
\bigskip
The Dyson-Wick perturbation theory expansion on the LF can be realized in
momentum space
by employing the Fourier transform of the fields and the
propagators.
 Many of the rules of the Feynman diagrams, for example,
 the symmetry factor $1/2$ for gluon loop, a minus sign associated with
fermionic loops etc.,  are the same as those found in
the conventional covariant framework \cite{lang}.  There are some
differences:$\,\,$ for example,  the external quark line
 now carries a factor $\theta(p^{+})\sqrt {{m}/{p^{+}}}$ while the
external gluon line a factor $\theta(q^{+})/\sqrt{2q^{+}}\;$ or that
 the Lorentz
invariant phase space factor is now
$\int d^{2}p^{\perp}dp^{+}\,\theta(p^{+})/(2p^{+})$.  The
external gluon line carries the polarization vector
$E^{\mu (\perp)}(q)=E^{\mu}_{(\perp)} = -D^{\mu}_{\perp}(q)$.  Its properties
 and the sum over the two polarization states are given Section 3.
The notation for
the quark field is given in Appendix B.
 Some of the momentum space vertices, used in Section 5,
 are collected below and others can be derived easily.

\bigskip
\nl {\it Gluon propagator}:
$$
 i\delta^{ab} \frac {D_{\mu\nu}(q)}{q^{2}+ i\epsilon}, \quad
{\rm with} \quad \,
  D_{\mu\nu}(q)=
    \left( -g_{\mu\nu} + \frac{ n_\mu q_\nu + q_\mu n_\nu }{ n\cdot q }
    -\frac{ q^2  }{ (n\cdot q)^2}  n_\mu  n_\nu \right),
$$
where $q_\mu$ is the gluon 4-momentum and $n_\mu$ is the
gauge direction.  We choose $n_\mu\equiv \delta^{+}_\mu$ and
 $n^{*}_\mu\equiv \delta^{-}_\mu$, the dual of $n_\mu$.
The useful
properties of the projector $D_{\mu\nu}$ are given in Eqs. (16-19).

\bigskip
\nl {\it Quark propagator}:

$$
   i\delta_{ij} S(p),\quad {\rm with} \quad  S(p) \equiv
   \frac {\not {p} +m}{p^2-m^2+i\epsilon}-
    \frac{\gamma^{+}}{2 p^{+}}= \frac {N(p)}{p^2-m^2+i\epsilon}
\,\qquad \epsilon> 0,
$$
where $p_\mu$ is the quark 4-momentum, $i$ and $j$ are color
indices and $N(p)=({\not {p}}+m)$ $-
    (p^{2}-m^{2}){\gamma^{+}}/{2 p^{+}} $.
The noncovariant second term on the right hand side is present only in the
propagator of the nondynamical dependent field $\psi_{-}$.
Its contribution, for example,  to gluon self-energy (63) is, however,
 compensated by that arising from the fourth term in (24).

\bigskip
\nl {\it Quark-quark-gluon vertex factor}:

$$ ig\,\gamma^\mu t^a $$

\bigskip
\nl {\it 3-gluon vertex factor}:
$$
-g \,f_{abc} \,F_{\lambda\mu\nu}(p_{1},p_{2},p_{3})
$$
Here
 $(p_{1},{\lambda},a)$,
$(p_{2},{\mu},b)$, and $(p_{3},{\nu},c)$,
 label the three gluons at the vertex with outgoing momenta $p_{1}^{\lambda}$,
 $p_{2}^{\mu}$ and $p_{3}^{\nu}$ with the associated gauge indices $a$, $b$ and
$c$ respectively.  Also
$ F_{\lambda\mu\nu}(p_{1},p_{2},p_{3})= (p_{1}-p_{2})_{\nu}g_{\lambda\mu} +
(p_{2}-p_{3})_{\lambda}g_{\mu\nu}+(p_{3}-p_{1})_{\mu}g_{\nu\lambda}
=-F_{\lambda\nu\mu}(p_{1},p_{3},p_{2})$ and $p_1+p_2+p_3=0$.

\bigskip
\nl {\it 4-gluon vertex factors}:

There are two types of 4-gluon vertices in this framework.
The ${V_{1}\,}^{abcd}_{\alpha\mu\nu\beta}$ is the usual momentum-independent
four-gluon vertex of the covariant gauge theory, while
${V_{2}\,}^{abcd}_{\alpha\mu\nu\beta}(p_1,p_2,p_3,p_4)$ is a new vertex
generated by
l.c. gauge LF quantized QCD,
with the momentum dependence as well, coming from the last term in
the interaction Hamiltonian (36):
\begin{eqnarray}
{V_{1}\,}^{abcd}_{\alpha\mu\nu\beta}&=& -ig^2 \left[
  f^{eab}f^{ecd} (g^{\alpha\nu}g^{\beta\mu}-g^{\alpha\beta}g^{\mu\nu}) \right.
\nonumber\\
&& \left. + f^{eac}f^{ebd}
(g^{\alpha\mu}g^{\beta\nu}-g^{\alpha\beta}g^{\mu\nu})
\right. \nonumber\\
&& \left. + f^{ead}f^{ecb}(g^{\alpha\nu}g^{\beta\mu}-g^{\alpha\mu}g^{\beta\nu})
  \right] \nonumber
\end{eqnarray}
and
\begin{eqnarray}
{V_{2}\,}^{abcd}_{\alpha\mu\nu\beta}(p_1,p_2,p_3,p_4)&=& +ig^2 \left[
  f^{eab}f^{ecd} g^{\alpha\mu}g^{\beta\nu}\, \frac
{(p_1-p_2)^{+}(p_3-p_4)^{+}}{(p_1+p_2)^{+ 2}} \right.
\nonumber\\
&& \left. + f^{eac}f^{ebd} g^{\alpha\nu}g^{\beta\mu}\, \frac
{(p_1-p_3)^{+}(p_2-p_4)^{+}}{(p_1+p_3)^{+ 2}} \right.
\nonumber \\
&& \left. + f^{ead}f^{ecb} g^{\alpha\beta}g^{\mu\nu}\, \frac
{(p_1-p_4)^{+}(p_3-p_2)^{+}}{(p_1+p_4)^{+ 2}} \right] \nonumber
\end{eqnarray}
Here $(p_1,\alpha,a)$,  $(p_2,\mu,b)$,  $(p_3,\nu,c)$, and
 $(p_4,\beta,d)$ indicate the four outgoing gluons at the vertex
with $p_1+p_2+p_3+p_4=0$.

\bigskip

\begin{center}{\Large \bf Appendix B}  \end{center}

\bigskip

\nl {\bf Spinor Field on the LF}

\medskip

The notation for gamma matrices is as given in Bjorken and Drell and
$\,\gamma^{\pm}=(\gamma^{0}\pm \gamma^{3})/{\sqrt
{2}}\,$ which satisfy
$\,({\gamma^{+}})^{2}=({\gamma^{-}})^{2}=0$.  The
$\,\Lambda^{\pm}= {1\over 2} \gamma^{\mp}\gamma^{\pm}=
{1\over \sqrt {2}} \gamma^{0}\gamma^{\pm}\,$ are hermitian projection
operators.  The spinor field on the
LF is decomposed naturally into $\,{\pm}\,$ projections:
$\,\psi_{\pm}=\Lambda^{\pm}\psi\,$,
$\,{\bar\psi}_{\pm}={\bar\psi}\Lambda^{\mp} \,$,
 $\psi=\psi_{+}+\psi_{-}$,
 and $ {\bar\psi} =\psi^{\dag}\gamma^{0}$
 $= {\bar\psi}_{+}+{\bar\psi}_{-}$,
$\,\gamma^{\pm}\psi_{\mp}=0 $,  $\,{\bar\psi}_{\pm} \gamma^{\mp}=0$,
{\em etc}.

The Fourier transform of the free spinor field \cite{cov} is
$$\psi(x)={1\over {\sqrt {(2\pi)^{3}}}} \sum_{r={\pm}}\int d^{2}p^{\perp}
         dp^{+}\,        \theta(p^{+}){\sqrt{m\over p^{+}}}\left[b^{(r)}(p){
u^{(r)}}(p)      e^{-ip\cdot x}+ d^{{\dag}{(r)}}(p){ v^{(r)}}(p) e^{ip\cdot
x}\right] $$
where
$$
u^{(r)}(p) = \frac{1}{({
{\sqrt {2}}p^{+} m })^{1\over 2}} \left[{\sqrt 2} p^{+}\Lambda^{+}
+(m+\gamma^{\perp}p_{\perp}) \Lambda^{-}\right]
\widetilde u^{(r)} \nonumber
$$
and the constant spinors $\widetilde u^{(r)}$ satisfy
$\gamma^{0}\widetilde u^{(r)}=\widetilde u^{(r)}$ and $ \Sigma_{3}
\widetilde
u^{(r)}= r \widetilde u^{(r)}$ with $\Sigma_{3}= i\gamma^{1}\gamma^{2}$ and
$r=\pm $.  The normalization and the completeness relations are:
$$\bar u^{(r)}(p)u^{(s)}(p)=\delta_{rs}=-\bar v^{(r)}(p)v^{(s)}(p)$$
$$\sum_{r=\pm} u^{(r)}(p){\bar u}^{(r)}(p)= \frac{(\not p+m)}{ {2m}}$$
$$\sum_{r=\pm} v^{(r)}(p){\bar v}^{(r)}(p)= \frac{(\not p-m)} {{2m}}$$
and C is the charge conjugation matrix \cite{bjk} while
$v^{(r)}(p)\equiv
{u^{(r)}(p)_{c}}= C {\gamma^{0}}^{T} u^{(r)}(p)^{*}$.

The $\Lambda^{+} $
projection of is by construction very simple,
${u^{(r)}}_{+}(p)={({\sqrt {2}}p^{+}/m)}^{1\over 2} (\Lambda^{+}{\widetilde
u^{(r)}})$.  The ${u^{(r)}}_{+}(p)$ are eigenstates of $\Sigma_{3}$ as
well, while the $\widetilde u^{(r)}$ correspond to rest frame spinors
when ${\sqrt 2}p^{\pm}=m$.

The free propagator for the independent component $\psi_{+}$ is
$$
\VEV{0|T({\psi^{i}}_{+}(x){\psi^{\dag
j}}_{+}(0))|0} = {{i\delta^{ij}}\over {(2\pi)^{4}}} \int d^{4}q
\;{{{\sqrt {2}}q^{+}\, \Lambda^{+} }\over {(q^2-m^{2}+i\epsilon)}}\,
e^{-iq.x}$$

\nl and we recall that $2i\partial_{-}\psi_{-}=(i\gamma^{\perp}\partial_{\perp}
+m)\gamma^{+}\psi_{+}$.

A detailed discussion of the properties of Dirac,
Majorana and Weyl
spinor fields,  helicity and spin operators, and
the generalized chiral invariance
of massive Dirac equation on the LF may be found in Ref. \cite{pre}.

\bigskip

\begin{center}{\Large \bf Appendix C}  \end{center}

\bigskip

The self-consistent canonical quantization of the
gauge theory requires one to fix the gauge and add ghost fields in
order to ensure BRS symmetry.  One must also
choose a
regularization procedure in order to compute and manipulate
the divergent integrals present in the
momentum space representation of the theory.
In our context
dimensional regularization is convenient
since it preserves the gauge (and BRS) symmetry, thus making the
task of renormalization simpler.  However the
regularization procedure
may break essential properties of a physical theory,
such as unitarity and introduce ghost degrees of freedom at an
intermediate stage.
One hopes that at the end of computations when
the regularization is removed
that we recover the features associated with the physical theory
represented by the original Lagrangian.
It is worth recalling the case of
Chiral Schwinger model where the regularization ambiguity itself
leaves behind an arbitrary parameter in the theory,  which is
physical and unitary only for a certain range of the
values of that parameter.  It was also shown a long time ago \cite{Gupta}
that the Pauli-Villars method of regularization corresponds to the
introduction of (regularization-dependent) ghost degrees of freedom in
the original theory.
The dimensional regularization procedure along with the
 ML prescription
may also give rise to the additional
(hidden regularization-dependent) ghost contributions to the divergent
integrals \cite{Morara:1999zu}. 
The noncovariant 
 $\ln (-2 q^+q^-) $ contribution in Eq. (42) to
the gluon self-energy has
an unconventional branch cut when compared to the normal threshold 
branch point in Eq. (39).  
It is an indication, in the
dimensional regularization scheme with ML prescription,  of ghost states
that do not propagate in the transverse direction.  The sign of the
 cut in Eq. (42) is that of asymptotic freedom, {\sl  i.e.},
ghostly.  In the gluon self-energy the sign of the
coefficient of $\ln (-q^{2}) $ term
in Eq. (39) is consistent with the spectral representation of
 K\"allen-Lehmann but that of $\ln (-2 q^{+}q^{-}) $ term is opposite.
 In the renormalization of the vertex functions, only the $\ln(-q^{2})$ 
terms are found.

We note that the renormalization factors of Yang-Mills theory in
Feynman gauge employing dimensional
regularization are
$$Z_{3}=1+\frac{g^{2}}{16\pi^{2}}C_{A}\, \frac{10}{3} \,
\frac{1}{\epsilon}\; >1, \qquad\qquad Z_{1}=
1+\frac{g^{2}}{16\pi^{2}}C_{A}\, \frac{4}{3} \, \frac{1}{\epsilon}\; >1, $$
\nl while
$${\tilde Z}_{3}=1+\frac{g^{2}}{16\pi^{2}}C_{A} \,
\frac{1}{\epsilon}\; >1, \qquad \qquad
{\tilde Z}_{1}=1-\frac{g^{2}}{16\pi^{2}}C_{A}\,
\frac{1}{\epsilon}\; < 1, $$
\nl such that $Z_{1}/Z_{3}={\tilde Z}_{1}/{\tilde Z}_{3}.$
The result is the same coupling constant renormalization as in our
LF framework
$$Z_{g}=\frac{Z_{1}}{{Z_{3}}^{3/2}}= \frac{{\tilde Z}_{1}}
{{\tilde Z}_{3}\sqrt{Z_{3}}}=
1-\frac{g^{2}}{16\pi^{2}}C_{A}\, (\frac{11}{6})\, N_{\epsilon}\; < 1. $$
\noindent Here ${\tilde Z}$ refer to the Faddeev-Popov ghost fields of
the covariant gauge theory.  We find that in this framework
 $Z_{3} > 1$, just as in the LF framework presented where, in
addition,
$Z_{g}={Z_{3}}^{-1/2}$.  This corresponds to the asymptotic freedom
property of dimensionally regularized nonabelian gauge theories
due to the three gluon's nonlinear self-interaction.

If only physical degrees of freedom are present, wavefunction
renormalization requires $Z_{3} < 1$; i.e, the probability that the
gluon is a bare gluon must be less than one.
As pointed out by Thorn \cite{thorn}, the fact that $Z_{3} > 1$ in
Yang-Mills theory is in conflict with the unitarity and the positivity of
the K\"allen-Lehmann spectral representation of the gluon propagator.
Note that in our light-front formulation of the
doubly transverse light-cone gauge,  the renormalization of the
three-gluon vertex receives a crucial one-loop contribution from the
instantaneous $V_2$ interaction as illustrated by the swordfish diagram
of figure 2(c). [This diagram is missing from the renormalization of the
proper three-gluon vertex in the conventional light-cone gauge formalism
since it is classified as one-particle reducible.] Similarly, the
tadpole graph 1(c) from $V_2$ must be included in computation of the
gluon self energy.  Such a term does not have a conventional two-particle
cut or K\"allen-Lehmann dispersion representation, thus allowing for
$Z^g_3 > 1$ \cite{Hwang}.
We note also that our gauge propagator is not diagonal like that in the
Feynman gauge.

\section*{Acknowledgments}


Comments from Robert Delbourgo, Valentin Franke,
John Hiller, Dae Sung Hwang, George Leibbrandt, Gary McCartor,
Sergey Paston, Olivier Piguet, Evgeni Prokhvatilov, Roberto Soldati,
Silvio Sorella, and Charles Thorn are thankfully acknowledged.  The
hospitality offered to PPS at the Theory Division of SLAC and financial
grants from CNPq and FAPERJ of Brazil for his participation at the
ICHEP2000, Osaka, Japan, are also gratefully acknowledged.

\newpage

\end{document}